# PACKMOL- GUI: An All-in-One VMD Interface for Efficient Molecular Packing

Jian Huang[†,1,2,3,5], Chenchen Wu[†,6], Xiner Yang[4], Zaixing Yang[4], Shengtang Liu[*,4], and Gang Yu[*,1,2,3,5]

1 Department of Data and Information, The Children's Hospital Zhejiang University School of Medicine, Hangzhou 310052, China;
2 Sino-Finland Joint AI Laboratory for Child Health of Zhejiang Province, Hangzhou 310052, China;
3 National Clinical Research Center for Child Health, Hangzhou 310052, China;
4 State Key Laboratory of Radiation Medicine and Protection, School for Radiological and Interdisciplinary Sciences (RAD-X), Collaborative Innovation Center of Radiation Medicine of Jiangsu Higher Education Institutions, Soochow University, Suzhou 215123, China;
5 Polytechnic Institute, Zhejiang University, Hangzhou 310052, China
6 Department of Radiation Oncology, The First Affiliated Hospital of Soochow University, Suzhou 215006, China

† These authors contributed equally.

* Corresponding authors: liushengtang@suda.edu.cn (Shengtang Liu), yugbme@zju.edu.cn (Gang Yu)

***Abstract:*** PACKMOL is a widely utilized molecular modeling tool within the computational chemistry community. However, its perceivable advantages have been impeded by the long-standing lack of a robust open-source graphical user interface (GUI) that integrates parameter settings with the visualization of molecular and geometric constraints. To address this limitation, we have developed PACKMOL-GUI, a VMD plugin that leverages the dynamic extensibility of Tcl/Tk toolkit. This GUI enables the configuration of all PACKMOL parameters through an intuitive user panel, while also facilitating the visualization of molecular structures and geometric constraints, including cubes, boxes and spheres, among others via the VMD software. The seamless interaction between VMD and PACKMOL provides an intuitive and efficient all-in-one platform for the packing of complex molecular systems.

# INTRODUCTION

Molecular dynamics (MD) simulations are crucial computational methodologies for investigating the thermodynamic and kinetic behaviors of complex molecular systems. The effective assembly of an appropriate initial configuration containing diverse molecular mixtures is a prerequisite in the MD simulation workflow. PACKMOL[1, 2] is a most widely utilized program to address this issue. However, its potential is limited by the lack of an intuitive graphical user interface (GUI) and molecular visualization functionalities. While GUI such as GEMS-Pack[3] and Atomistica.online[4] have made significant advancements, they still fall short in providing a convenient and robust interface for the majority of PACKMOL users. Julio C. G. Correia and colleagues developed GEMS-Pack utilizing Python 2.7 and PyQt5. However, to the best of our knowledge, GEMS-Pack is currently inaccessible as a subprogram within a larger project. Moreover, the end of updates for Python 2.7 and the expected discontinuation of PyQt5 support will likely to pose challenges for users during installation. Stevan Armaković and team developed Atomistica.online, a web-based GUI for various computational chemistry applications, including PACKMOL (https://atomistica-online-packmol.anvil.app). However, it has limited PACKMOL parameter settings, no visualization for molecular and geometric constraints, and restricts calculations to one minute due to resource limitations. Thus, PACKMOL still lacks a friendly open-source GUI that integrates both molecular visualization and constraint visualization capabilities.

Visual Molecular Dynamics[5] (VMD) is a widely employed frontend tool developed using the C++ programming language in conjunction with the OpenGL library, offering capabilities for modeling, visualization, and processing MD trajectories. Notably, VMD also provides a Tcl/Tk interface. The inherent dynamic extensibility of Tcl/Tk permits developers to integrate new GUI plugins without necessitating modifications to the VMD source code or recompilation of the entire application. These plugins substantially mitigate the complexity associated with utilizing diverse backend computational software[6-10], thereby facilitating a more concise computational workflow. For instance, to alleviate the complexity associated with molecular manipulation command-line procedures in VMD, we have developed Molcontroller[11], a tool designed to streamline repetitive tasks such as translation, rotation, and merging of molecules. In this study, we designed an out-of-the-box plugin named "PACKMOL-GUI" within the VMD program to address the absence of a graphical interface in PACKMOL. This GUI provides a visual representation of molecular constraints, assists in the generation of input files via user panels, and subsequently invokes PACKMOL to achieve the desired outcomes. The development of PACKMOL-GUI is anticipated to expand the user base of both VMD and PACKMOL, while simultaneously improving the efficiency of constructing complex molecular systems.

## ■ INSTALLATION GUIDE

Since VMD is compatible with Windows, Linux, and macOS, PACKMOL-GUI inherently supports these operating systems. Users are required to download the files at https://github.com/VMD-Plugin/PACKMOL-GUI and place them within the "plugins/noarch/tcl" subdirectory of the VMD installation directory. To initialize PACKMOL-GUI, one must incorporate the "vmd_install_extension" command into the VMD startup file (vmd.rc or .vmdrc), thereby completing the setup process. The step-by-step installation instructions are available in the README file, which is accessible online.

## ■ THE WORKFLOW OF PACKMOL-GUI

The workflow of PACKMOL-GUI (**Figure 1**) is structured according to the data flow inherent in the PACKMOL program. Upon promptly configuring the PATH to invoke PACKMOL-GUI within VMD's startup file, the GUI for PACKMOL becomes accessible under the "Extensions/Modeling" submenu within the main menu of VMD. Users are required to specify the installation path of PACKMOL for the first time. This essential parameter, along with other settings such as tolerance, file type, and the working directory for the output file, is preserved in a file entitled "packmol_info.json", thereby enhancing user experience and promoting program configurability. After configuring these basic parameters, users can import molecules prepared for modeling via the molecular import module. Additionally, we provide users with access to a comprehensive collection of shared molecular structure datasets. These datasets are systematically organized to substantially improve users' efficiency in expeditious modeling. After importing molecules, users can employ the visualization capabilities of both VMD and PACKMOL-GUI to meticulously design the spatial arrangement and geometric constraints of the molecules. Following the establishment of these constraints and the setting of parameters, users can proceed to generate the requisite input files for PACKMOL. Ultimately, it is necessary to call PACKMOL and await the completion of the program's execution, which will result in the generation of the final output files.

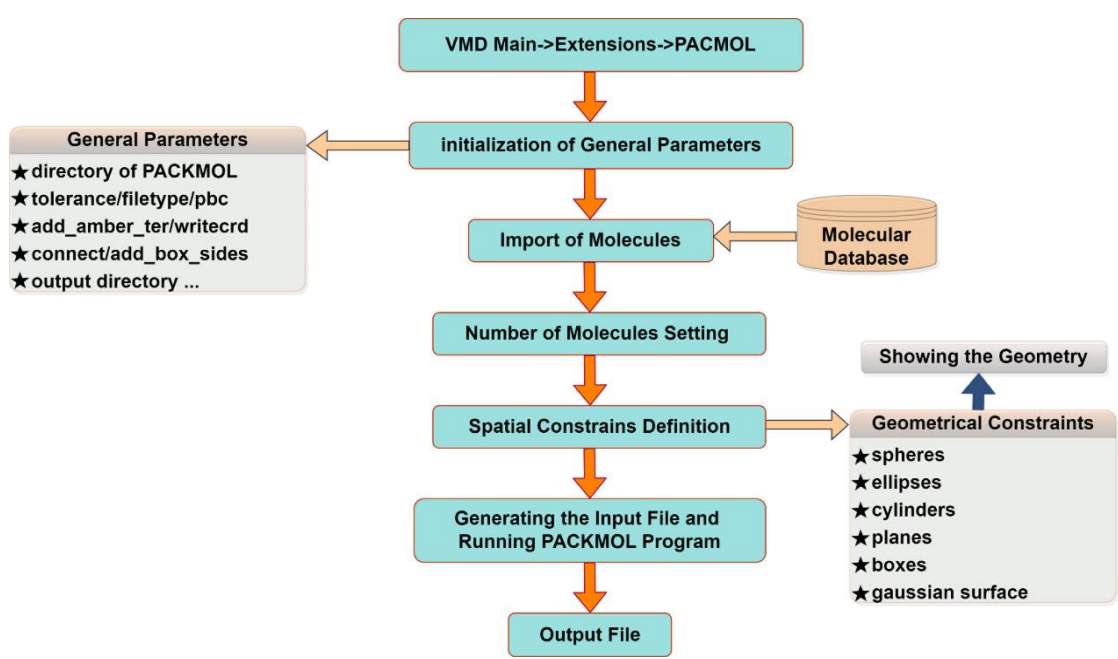

**Figure 1.** Overview of PACKMOL-GU workflow.

## ■ MAIN FEATURES

PACKMOL-GUI is designed with the same uniformity and simplicity as VMD-GUI. It allows users to set all PACKMOL parameters directly within the user panels, enhancing modeling efficiency through its intuitive "what you see is what you get" approach. Based on the functionalities of the PACKMOL program, we have structured the PACKMOL-GUI into five modules: "General Parameters," "Molecule Import," "Geometric Constraints," "Input File Generation and Execution," and "Output Results." **Figure 2** illustrates the PACKMOL-GUI, and the principal features of each module are described in the subsequent sections.

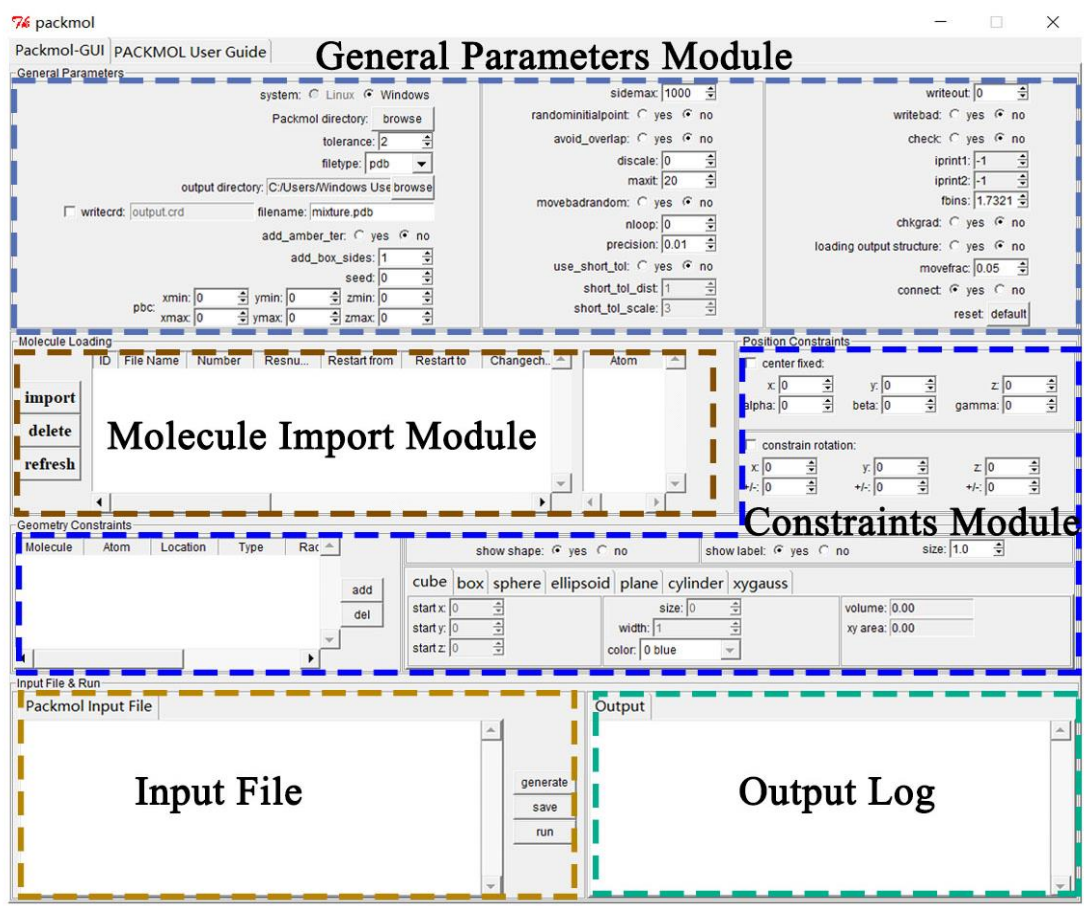

**Figure 2. Layout of the PACKMOL-GUI.** The layout of various modules within PACKMOL-GUI is systematically organized in a top-to-bottom sequence, comprising the General Parameters Module, Molecule Import Module, Constraints Module, and the Input File and Output Log Module. Each module is distinctly delineated by colored dashed borders.

**General Parameters Module.** When initialized, PACKMOL-GUI autonomously detects and adjusts to file directory formats in accordance with the user's operating system. Except for setting the PACKMOL executable path and output directory, the GUI will assign default parameters within the user panels of this module, such as file type being set to PDB format, a tolerance of 2.0, a seed value of -1, which generates the seed based on the computer's current time, among others. Additionally, an input field for defining periodic boundary conditions (PBC) has been integrated to ensure compatibility with the most recent version of PACKMOL (version 20.15.0, released on July 13, 2024). As previously indicated, these general parameters will be documented in a JSON-formatted file named "packmol_info.json." For detailed explanations of these parameters, users are advised to consult the PACKMOL user manual. Additionally, the contents of the manual have been integrated into the "PACKMOL User Guide" tab on the right side of the parameter panel, providing users with convenient access.

**Molecule Import Module.** This module encompasses three functional buttons, a compilation of molecular properties, and a catalog of atom names and indices. The buttons labeled "Import," "Delete," and "Refresh" execute their designated functions as follows: importing molecules, deleting molecules, and synchronizing the GUI's molecular list with the deletions performed in the VMD main window. The molecular properties list includes all PACKMOL parameters related to molecular characteristics, such as molecule numbers and residue numbering in PDB files. Users can view these properties by dragging the corresponding slider. For an in-depth understanding of these parameters, it is advisable to refer to the PACKMOL user guide. Considering the extensive attention of biomolecular and nanomaterial systems within the MD simulation community, we have integrated an dataset encompassing commonly utilized biomolecules, solution molecules, gas molecules, ions, and nanomaterials. Biomolecules are predominantly composed of various lipids and their derivatives, amino acids, and nucleotide bases. Solutions include ethanol, dimethyl sulfoxide (DMSO), and various models of water molecules. The category of ions encompasses common salt ions as well as radioactive nuclide ions. Gas molecules include typical atmospheric constituents and noble gases. Nanomaterials are primarily constituted of computation-ready experimental metal-organic frameworks[12] (CoRE-MOFs). These datasets offers a convenient platform for the modeling of various complex systems. After importing molecules, pertinent PACKMOL parameters, including residue ordering and chain identifiers, can be set up within the item. A critical parameter in PACKMOL is the number of molecules specified under geometric constraints. To evaluate the maximum number of molecules that can be accommodated within the constrained geometric space, we utilize two estimation methods based on volume and surface area. The formula for volume-based estimation is as follows:

$$N_{vmax} = \frac{V_{constraints}}{V_{molecule}}, \qquad (1)$$

where $V_{constraints}$ represents the volume of the constrained shape and $V_{molecule}$ denotes the volume of the individual molecule. In the context of practical applications involving MOFs as substrate interfaces for single molecule, we have calculated the molecular volumes of lipids and their derivatives, amino acids, and nucleotide bases.

For lipids within membrane systems, an surface-area-based estimation method was utilized. The specific formulas employed are as follows:

$$N_{smax} = \frac{S_{constraints}}{APL_{molecule}}, \quad (2)$$

where $S_{constraints}$ represents the membrane surface area of the constrained shape and $APL_{molecule}$ denotes the area per lipid (APL), indicating the space occupied by individual lipid molecules at the membrane-water interface. We document the calculated molecular volumes and APL values utilizing the REMARK 990 tag within the PDB file. These self-defined parameters are also present within the list of molecular properties. For custom molecules, users can efficiently compute the molecular volumes employing tools such as MoloVol program[13] to estimate the maximum number of molecules. Once the missing constrained volumes (or membrane surface areas) and molecular volumes (or APL) are added to the lists, $N_{vmax}$ (or $N_{smax}$) will be automatically generated. Furthermore, loaded molecules can be readily distinguished in VMD by employing various rendering modes, thereby enhancing the intuitive understanding of subsequent spatial and geometric constraint definitions. We have also incorporated a default highlighting mode within the selected atomic subset of the molecule, wherein the rendering method is configured to "Licorice" and the rendering color is assigned as yellow in VMD, to accentuate the constrained atoms.

**Constraints Module.** The constraint functionality represents the most distinctive feature of the PACKMOL program, encompassing position constraints, rotational constraints, and most notably geometric constraints. In this module, users are afforded the capability to add, modify, or delete constraints pertaining to molecules or their constituent atoms. The "Location" label provides a dropdown menu with options such as "inside," "outside," "over," and "below." Concurrently, the "Type" label encompasses a variety of geometric forms, including cube, box, sphere, ellipsoid, plane, cylinder, and xygauss (Gaussian surface). Restrictions are imposed on these labels: "inside" and "outside" are applicable exclusively to cube, box, sphere, ellipsoid, and cylinder geometries, whereas "over" and "below" are pertinent solely to plane and Gaussian surface geometries. The geometric shapes and their corresponding coordinate points (e.g. center of the sphere) are visualized in the VMD display window following the specification of parameters and the activation of the Enter key. Users can enable or disable these shapes and coordinate labels using the "show shape" and "show label" radio button located at the top of the geometry panel. Additionally, users can modify the line thickness, color, and material properties of the shapes and text size of labels within the panels to enhance visual representation. By employing VMD's molecular measurement tools, the Molcontroller, and PACKMOL-GUI, users can effectively position molecules within geometric constraints and precisely estimate shape parameters.

**Input File Generation and Execution Module.** After configuring all parameters, users can select the "generate" button to produce PACKMOL input files, which will be displayed in the text box on the left. A "save" button is provided for users to store these files in a specified location. To mitigate the risk of users neglecting to save or inadvertently losing files, PACKMOL-GUI autonomously generates input files within the working directory, appending a timestamp upon activation of the "generate" button. After configuring all settings and verifying the parameters, users can initiate the PACKMOL program by selecting the "run" button, thereby executing the process in the backend.

**Output Results Module.** To facilitate user monitoring of the PACKMOL execution status, the program's output information is redirected to the text box of the output module. This functionality allows users to efficiently identify, locate, and rectify errors of input file.

## ■ CASE STUDIES

In order to showcase the promising performance of PACKMOL-GUI, we used the "Double layered spherical vesicle with water inside and outside" example provided on the PACKMOL website (https://m3g.github.io/packmol/examples.shtml) and a radionuclide ions exchange system within a cationic MOFs material from our prior research[14].

**Case 1: Palmitoyl Spherical Vesicle**

The palmitoyl vesicle comprises a spherical aqueous core with a radius of 13 Å, surrounded by a bilayer palmitoyl membrane, and an external water box with a side length of 95 Å. Initially, the palmitoyl.pdb file was obtained from the PACKMOL website and placed in the working directory for subsequent modeling. The analyses suggest that this mixed system imposes four distinct types of geometric constraints on water molecules and palmitoyl. Consequently, palmitoyl and TIP3P water molecules (sourced from the dataset/water directory) were imported twice, resulting in four molecules arranged from the interior to the exterior of the vesicle: water-0, palmitoyl-1, palmitoyl-2, and water-3 (**Figure 3a**). For the water-0, we implemented an "inside sphere" constraint by specifying the sphere's center coordinates as (0, 0, 0) and a radius of 13 Å in the parameter panel of sphere. Upon confirmation by pressing Enter key, a default blue sphere with a radius of 13 Å was immediately rendered in the VMD display window. Next, users can utilize VMD's measurement tools to determine the length of palmitoyl, which is approximately 12 Å. Given that the hydrophilic carboxyl head (atom numbering 37) of palmitoyl orients towards the aqueous environment, while the hydrophobic methyl tail (atom numbering 5) aggregates with the hydrophobic tail of another palmitoyl molecule, it follows that the hydrophilic head of palmitoyl-1 is limited to a 14 Å radius sphere. In contrast, the hydrophobic tail is confined outside a spherical region with a radius of 26 Å (14 Å + 12 Å). In a similar manner, the hydrophobic tail of outer membrane palmitoyl-2 is confined within a sphere of 29 Å radius, whereas the hydrophilic head is confined outside a sphere of 41 Å radius. Finally, the vesicle is solvated within a water box possessing a side length of 90 Å. Consequently, the final constraint is applied to water-3, which is confined within a cubic box centered at the origin (0,0,0), with boundary coordinates extending from (-45, -45, -45) to (45, 45, 45). In addition, water-3 must also satisfy the geometric constraint of being positioned outside a sphere centered at the origin (0, 0, 0) with a radius of 43 Å, which incorporates an additional tolerance of 2 Å beyond the 41 Å vesicle. The quantity of molecules for each constraint type is determined based on the PACKMOL website example. Notably, the maximum number of molecules determined through the GUI has been observed to converge. Users can seamlessly enter these general parameters, along with geometric constraint parameters, into the designated fields within our GUI panel to generate the corresponding PACKMOL input file. During the construction of these geometric constraints, users can dynamically reposition water-0, palmitoyl-1, palmitoyl-2, and water-3 to their respective geometries utilizing Molcontroller. By modifying the material properties and transparency, the geometries can be rendered with clarity, thereby enabling users to verify the spatial constraints for accuracy. The geometric constraints employed throughout the construction process are depicted in **Figure 3**.

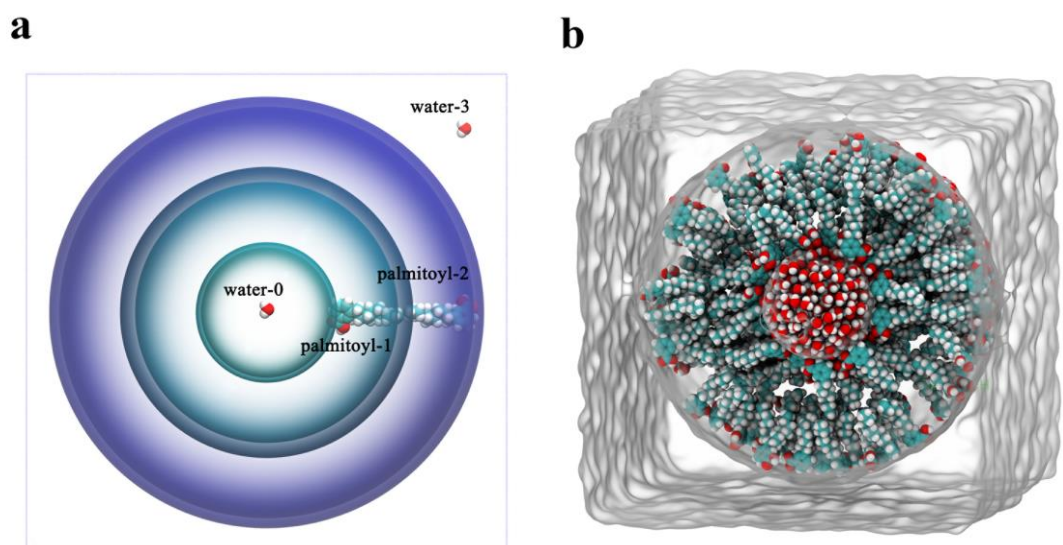

**Figure 3. "Double layered spherical vesicle with water inside and outside" example**. **a)** Four distinct spatial geometries were employed as constraints for water and palmitoyl molecules. The interior aqueous core and interior membrane was represented by cyan GlassBubble, whereas the exterior membrane was depicted using blue GlassBubble. These molecules were moved within their respective geometries utilizing the Molcontroller tool. **b)** The initial configuration illustrates the solvent box rendered with a silver QuickSurf, while the vesicle and the interior aqueous core are displayed in cross-section. Molecules are drawn in vdW mode, with carbon shown in cyan, hydrogen in white, and oxygen in red.

**Case 2: SCU-103 Enrichment of $^{99}TcO_4^-$**

SCU-103 is a cationic metal-organic framework material featuring $Ni_2^+$ as the metal node, employed for the enrichment of $^{99}TcO_4^-$ in competitive anion environments, including $OH^-$, $NO_3^-$, and $SO_4^{2-}$. In our prior research, the initial configuration of this system was constructed iteratively using tools such as GROMACS insert-molecules and Molcontroller, a process that proved to be both cumbersome and time-consuming. In this study, we utilized PACKMOL-GUI to reconstruct the SCU-103 system, which comprises $3 \times 3 \times 1$ unit cells with lattice parameters of 55.30 Å $\times$ 47.90 Å $\times$ 8.20 Å, and contains 24 $NO_3^-$ counterions. To simulate a complex competitive anion environment, we distributed 12 $OH^-$, 12 $NO_3^-$, 12 $SO_4^{2-}$, and 12 $^{99}TcO_4^-$ evenly on each side of the SCU-103 surface at a distance of 12 Å, which corresponds to the nonbonded interaction cut-off distance in MD simulations. Additionally, to replicate the experimental conditions, 24 $NH_3^+$ and 96 $Na^+$ ions were incorporated to neutralize the system's charge. The final system is solvated by 11,118 water molecules within a box of dimensions 55.30 Å × 47.90 Å × 140.00 Å. Accordingly, the SCU-103 was placed at the origin, with box boundaries set at coordinates (-27.65, -23.95, -4.10) and (27.65, 23.95, 4.10). Furthermore, given that the initial positions of these ions are set 12 Å away from the SCU-103 surface, the upper and lower boundary values for these ions are -16.10 Å and 16.10 Å, respectively. By importing SCU-103, ions, and water molecules, we can efficiently impose spatial constraint and invoke PACKMOL to generate the initial configuration of the mixed system, as illustrated in **Figure 4**.

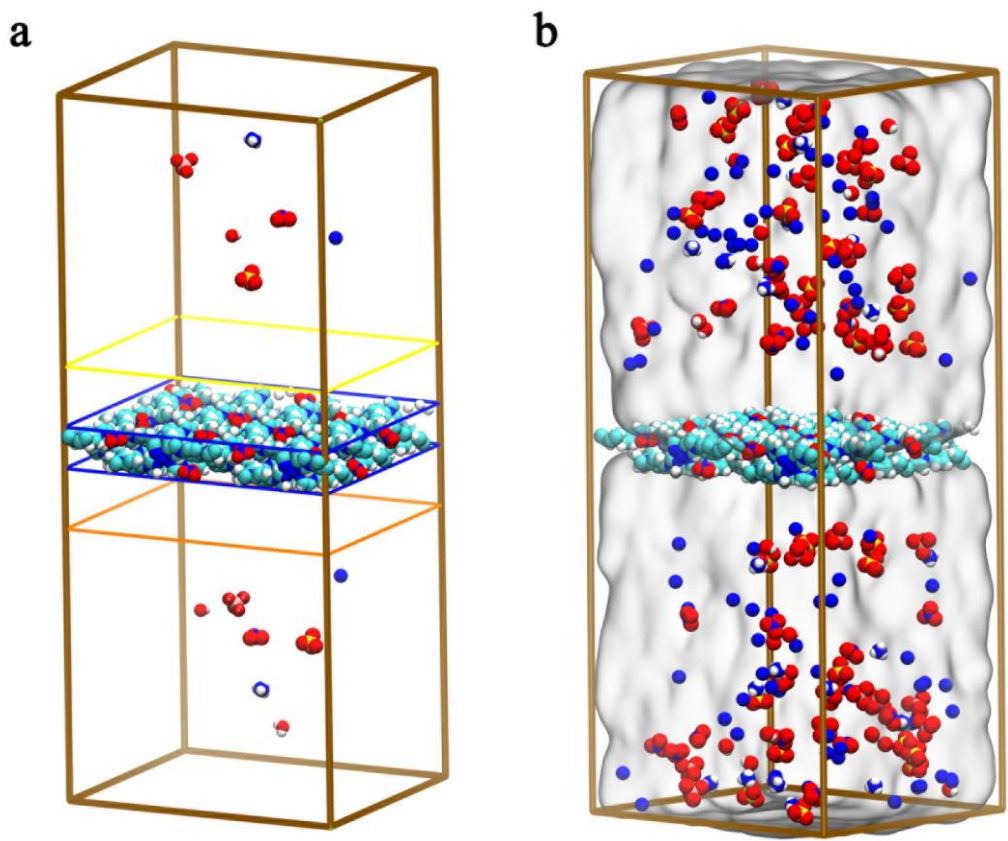

**Figure 4. Cationic MOFs SUC-103 are utilized for the adsorption of $^{99}TcO_4^-$. a)** The geometric constraints visualized using PACKMOL-GUI are illustrated. The SCU-103 constraint is represented by a blue box, while yellow and orange boxes delineate the regions proximal to SCU-103 where various ions are confined. The water solvent box is depicted in ochre. All ions are depicted using the vdW representation, with $^{99}Tc$ illustrated in orange, sulfur atoms in yellow, oxygen atoms in red, hydrogen atoms in white, and nitrogen atoms in blue. $Na^+$ are indicated by small blue spheres. **b)** The initial configuration generated for MD simulations is depicted, with the water solvent box

represented using a silver QuickSurf.

## ■ TECHNICAL IMPLEMENTATION

The PACKMOL program requires the latest version of 20.15.0. The PACKMOL-GUI (https://github.com/VMD-Plugin/PACKMOL-GUI) and Molcontroller (https://github.com/VMD-Plugin/Molcontroller) are recommended for use with VMD version 1.9.3. All lipids and their derivative molecules, along with the associated APL values, were sourced from the CHARMM-GUI[15-19]. Molecular volumes were computed using the MoloVol software[13], employing van der Waals parameters as described by Santiago Alvarez[20]. The molecular volume of water was calculated based on its density and molar mass under standard conditions, resulting in a value of 29.9 Å³. All CIF files from the CoRE-MOFs database were converted to PDB format using OpenBabel[21], and their unit cells were expanded to dimensions exceeding 24 Å to meet the requirements of most molecular modelling studies.

## ■ CONCLUSION

The PACKMOL program is favored within the computational chemistry community for its extensive geometric constraints and reliable convergence. Nevertheless, its full potential remains underutilized due to the lack of an robust open-source graphical user interface. To compensate for this shortcoming, we exploit the dynamic extensibility of Tcl/Tk scripting capabilities to develop the PACKMOL-GUI for visualizing PACKMOL's parameter panels, molecular structures, and geometric constraints. The integration of PACKMOL-GUI within the VMD program provides a streamlined workflow for users of both VMD and PACKMOL, significantly enhancing the efficiency of molecular modeling for large-scale systems within the community.

**AUTHOR INFORMATION**

**Corresponding Authors**

**Gang Yu** − *Department of Data and Information, The Children's Hospital Zhejiang University School of Medicine, Hangzhou 310052, China; Sino-Finland Joint AI Laboratory for Child Health of Zhejiang Province, Hangzhou 310052, China; National*

*Clinical Research Center for Child Health, Hangzhou 310052, China; Polytechnic Institute, Zhejiang University, Hangzhou 310052, China*; orcid.org/0000-0001-9935-9969; Email: yugbme@zju.edu.cn

**Shengtang Liu** − *State Key Laboratory of Radiation Medicine and Protection, School for Radiological and Interdisciplinary Sciences (RAD-X) and Collaborative Innovation Center of Radiation Medicine of Jiangsu Higher Education Institutions, Soochow University, Suzhou 215123, China*; orcid.org/0000-0001-9632-9377; Email: liushengtang@suda.edu.cn

### ■ AUTHORS

**Jian Huang** − *Department of Data and Information, The Children's Hospital Zhejiang University School of Medicine, Hangzhou 310052, China; Sino-Finland Joint AI Laboratory for Child Health of Zhejiang Province, Hangzhou 310052, China; National Clinical Research Center for Child Health, Hangzhou 310052, China*; orcid.org/0000-0002-1955-4316

**Chenchen Wu** − *Department of Radiation Oncology, The First Affiliated Hospital of Soochow University, Suzhou 215006, China*

**Xiner Yang** − *State Key Laboratory of Radiation Medicine and Protection, School for Radiological and Interdisciplinary Sciences (RAD-X), Soochow University, Suzhou 215123, China*

**Zaixing Yang** - *State Key Laboratory of Radiation Medicine and Protection, School for Radiological and Interdisciplinary Sciences (RAD-X), Soochow University, Suzhou 215123, China*; orcid.org/0000-0003-3521-6867

### ■ CONTRIBUTIONS

S.L. and G.Y. conceived and designed the research. J.H. and C.W. developed the programs. X.Y. and S.L. conducted the validation. J.H., C.W., Z.Y., S.L., and G.Y. collaborated on writing the paper. All authors participated in discussions and provided feedback on the manuscript.

■ NOTES

The authors declare no competing financial interests.

■ ACKNOWLEDGMENT

This work was partially supported by the National Natural Science Foundation of China (grant numbers 22106114, 22176137 and 62076218), the National Key Research and Development Program of China (grant numbers and 2019YFE0126200), the Natural Science Foundation of the Jiangsu Higher Education Institutions of China (20KJA150010), the Priority Academic Program Development of Jiangsu Higher Education Institutions (PAPD), and the Jiangsu Provincial Key Laboratory of Radiation Medicine and Protection.